\documentclass[showpacs,preprintnumbers,amsmath,amssymb]{revtex4}
\usepackage{graphicx}
\begin{document}
\title{      Environmental  Effects  on the  Terahertz Surface Plasmons  in Epitaxial Graphene
}
\author{ Godfrey Gumbs$^{1}$, Andrii Iurov$^2$, Jhao-Ying Wu$^3$,  M. F. Lin$^3$, Paula Fekete$^4$ }
\affiliation{$^{1}$Department of Physics and Astronomy\\
Hunter College of the City University of New York\\
695 Park Avenue, New York, NY 10065 USA\\
$^2$ Center for High Technology Materials, University of New Mexico, Albuquerque, New Mexico, 87106, USA\\
$^3$ Department of Physics, National Cheng Kung University, Tainan, Taiwan 701\\
$^4$ US Military Academy, West Point, NY
}

\begin{abstract}

 In this paper, we predict the existence of low-frequency nonlocal plasmon excitations  at the
vacuum-surface interface of a  superlattice of $N$ graphene  layers interacting with
a thick conducting substrate. This is different from graphite  which allows
inter-layer hopping.
A dispersion function is derived which incorporates the polarization function of the graphene
monolayers (MLGs) and the dispersion function of a semi-infinite  electron liquid at whose
surface the  electrons scatter specularly. We find
that this surface plasmon-polariton is not  damped by the particle-hole
excitations (PHEs) or the bulk modes and separates below the continuum mini-band of bulk plasmon modes.
For a conducting substrate with surface plasmon frequency $\omega_s=\omega_p/\sqrt{2}$, the surface
plasmon frequency of the hybrid structure always lies below $\omega_s$.
The intensity of this mode depends on the distance of the graphene layers from the surface
of the conductor,  the energy band gap between the valence and conduction bands of MLG
and, most importantly, on the number of  two-dimensional (2D)  layers. Furthermore,
 the hybrid structure has no surface plasmon for a sufficiently large number
 ($N\stackrel{>}{\sim} 7$) of layers. The existence of two plasmons with different
dispersion relations indicates  that   quasiparticles with different group velocity
may coexist for various ranges of wavelength which is determined by the number of layers
in the superlattice.

\end{abstract}


\maketitle

\section{  Introduction}
\label{sec:1}

Several years ago, there was considerable activity in the study of the
electronic properties of layered semiconductor superlattices  \cite{Esaki,Giuliani,Jain,Mancini,Tselis}
leading to intriguing  transport and optical properties which are the result of quantum
 mechanical effects on the nanoscale.  Specifically, it was desirable to learn how a 
 periodic or quasi-periodic array \cite{Ali}  of two-dimensional electron liquid (2DEL) 
 layers,  would lead to a modification of  the response of the charge-density excitations 
 to an electromagnetic field.   Recently, there has been some investigation regarding
the Coulomb excitations of an  $N$-layered superlattice of free-standing
monolayer graphene layers (MGLs) \cite{Peeters1}.  However,  what was neglected in that study
is the interaction between epitaxial layers of graphene  and a conducting substrate
which may give rise to  composite plasmon-plasmon resonances. A low-frequency
mode emerges below the continuum of modes in the limit of a  large number of layers
and is associated with the air-exposed surface of the graphene-conductor substrate combination.
As a matter of fact, several authors  \cite{Nickel,Pol1,Pol2,Pol3,Pol4,1} demonstrated 
how the electronic response properties of graphene-metal  composites of Ru and Ni, for 
example,  are much different from free-standing structures. Furthermore,  these complex
 carbon/metal interfaces are  interesting because of  the unusual and   fundamental physics  regarding their
electronic   and magnetic properties at the 2D  interfaces.
Examples of these systems occur in intercalated graphite \cite{interface1}, incommensurate transition
metal/graphene  \cite{interface2} and carbide/graphene interfaces  \cite{interface3} .
Possible motivation for pursuing this area of research is the tunability of graphene plasmons by a substrate  
which is a promising emerging field of graphene-based plasmonics.

\medskip
\par

Here, we demonstrate how the intensity of the response of the surface plasmon arising
 when  $N$ graphene layers interact with a conducting  substrate may be adjusted
 by changing the layer-substrate separation, the energy band gap or the number of 2D
 layers.  It has been shown that such an energy gap may be
 produced by circularly polarized light \cite{Kibis,Andrii}. We employ our model to determine
 the plasmon excitation dispersion relation for a single layer on a
 conducting substrate and show how  our results may simulate the experimental data
 recently reported by Politano, et al. \cite{Pol1,Pol2,Pol3,Pol4,1}.  Additionally, this paper   investigates
 multi-layer graphene generally, showing how its new surface mode depends on the
 in-plane wave vector and the critical wave vector $q_c$ which marks the
 onset of  damping by bulk modes in the miniband continuum.

\medskip
\par

 The model we use for a superlattice consists of $N$ 2D graphene layers whose
 planes are perpendicular to the $z$-axis at $z=a_1,a_2,\cdots, a_N$ and the
 substrate occupies the half-space $z<0$. Each graphene layer will be described
 by an energy band structure for Dirac fermions and may be intrinsic, gapped
 or doped.    This model
does not allow inter-layer hopping and hence it differs from the one we employed  to describe the anisotropy of $\pi$-plasmon dispersion  in AA-stacked graphite  \cite{Chiu}.
The screening of an externally applied frequency-dependent
 potential by the polarized medium requires a knowledge of the dielectric
 function of the structure which we obtain in the random-phase-approximation
 (RPA).  We present our method of calculation in Sec. \ \ref{sec:2}.
Section \ref{sec:3} is devoted to numerical
results and discussion of our plasmon dispersion when several  graphene layers
interact with a conducting substrate. The simulated data show how the intensity
of the modes depends on the number of layers which are stacked, their distance
from the conducting surface as well as the energy gap. For $N=1$ gapless MLG, when
the conductor surface-MLG separation exceeds a critical distance $d_c$, the intensity
of the surface plasmon in the long wavelength regime is sufficiently high to be observable
up to some cut-off wave vector $q_c$ of the surface plasmon frequency. Beyond $q_c$, the intensity of $\omega_c$ is very weak until
the plasmon wave vector  reaches some  value $q_c^\prime$.
However, when  the surface-MLG separation is less  that  $d_c$, the surface plasmon
intensity in the long wavelength regime is weak and the mode only appears at shorter wavelengths
when the in-plane wave vector $q_\parallel>q_c^\prime$. Interestingly, for gapped graphene,
  $\omega_c$    is completely suppressed when  the surface-
layer separation  is less than $d_c$. We then make some concluding
remarks in Sec.\ \ref{sec:4} regarding the inspiration for our work,  a summary and
significance of our findings, and what new theoretical formalism is presented in our paper.

\medskip
\par

\section{General Formulation of the Problem}
\label{sec:2}

In our formalism, we consider a nano-scale system consisting
of an arbitrary number of 2D layers and a thick  conducting material.  The
layer may be monolayer graphene or a 2DEL such as a
semiconductor  inversion layer or HEMT (high electron mobility
transistor).  The graphene layer may have a gap, thereby
extending the flexibility of the composite system that
also incorporates a thick layer of dielectric material.
The excitation spectra
of allowable modes will be determined from a knowledge
of the non-local dielectric function
$ \epsilon ({\bf r},{\bf r}^\prime;\omega)$   which depends
on position coordinates ${\bf r}, {\bf r}^\prime$ and
frequency $\omega$ or its inverse
$ K({\bf r},{\bf r}^\prime;\omega)$ satisfying
$\int d{\bf r}^\prime \ K({\bf r},{\bf r}^\prime;\omega) \epsilon({\bf r}^\prime,{\bf r}^{\prime\prime};\omega)
=\delta({\bf r},{\bf r}^{\prime \prime})$.

In operator notation, the dielectric function  for the $N$ 2D layers and a semi-infinite
structure is given by \cite{Gumbs-PRB}

\begin{equation}
\hat{\epsilon}= \hat{1}+\hat{\alpha}_{SI}+\sum_{i=1}^N\hat{\alpha}_{2D}^{(i)}
 \equiv \hat{\epsilon}_{SI}+ \sum_{i=1}^N \hat{\alpha}_{2D}^{(i)}
=\hat{K}_{SI}^{-1}+  \sum_{i=1}^N\hat{\alpha}_{2D}^{(i)}  \ ,
\end{equation}
where $\hat{\epsilon}=\hat{K}^{-1}$ with $\hat{K}$ the inverse dielectric function satisfying

\begin{equation}
\hat{K}=\hat{K}_{SI} -\hat{K}_{SI} \cdot \sum_{i=1}^N  \hat{\alpha}_{2D}^{(i)} \cdot  \hat{K} \ .
\end{equation}
In integral form, after Fourier transforming parallel to the $xy$-plane and
suppressing the in-plane wave number $q_{||}$ and frequency $\omega$,
we obtain

\begin{equation}
K(z_1,z_2)= K_{SI}(z_1,z_2) - \sum_{i=1}^N \int_{-\infty}^\infty dz^\prime \int_{-\infty}^\infty
 dz^{\prime\prime}\  K(z_1,z^\prime)
\alpha_{2D}^{(i)}(z^\prime ,z^{\prime\prime})  K(z^{\prime\prime} ,z_2)  \ . 
\end{equation}
Here, the polarization function for the 2D structure is given by

\begin{equation}
\alpha_{2D}^{(i)}(z^{\prime}, z^{\prime\prime})= \int_{-\infty}^\infty  dz^{\prime\prime\prime} \
v(z^\prime, z^{\prime\prime\prime} ) D^{(i)}(z^{\prime\prime\prime},z^{\prime\prime})  \ ,
\end{equation}
where the 2D response function obeys

\begin{equation}
D^{(i)}(z^{\prime\prime\prime},z^{\prime\prime}) = \Pi_{2D,i}^{(0)} (q_{||},\omega)
  \delta(z^{\prime\prime\prime}-a_i) \delta(z^{\prime\prime}-a)
\end{equation}
with $\Pi_{2D,i}^{(0)} (q_{||},\omega)$  the single-particle in-plane
response \cite{wunch,pavlop,Shung2,Dsarma}. 
The 2D RPA ring diagram polarization function for graphene with a gap $\Delta$  may be expressed as

\begin{eqnarray}
\label{A1}
\Pi^{(0)}_{2D}(q_\parallel ,\omega)
&& = \frac{g}{4 \pi^2} \int d^2 {\bf k} \sum\limits_{s,s' = \pm} 
\left( 1 + s s' \frac{{\bf k} \cdot ({\bf k}+{\bf q}_{\parallel}) + \Delta^2}{\epsilon_k \,\, \epsilon_{\vert {\bf k}+{ \bf q}_{\parallel} \vert }}  \right) 
\nonumber \\ 
&& \times \frac{f(s \, \epsilon_{{\bf k}}) - f(s' \epsilon_{{\bf k}+{\bf q}_{\parallel}})}{s \, 
\epsilon_{{\bf k}_{\parallel}} - s' \epsilon_{{\bf k}+{\bf q}_{\parallel}} - \hbar \omega - i 0^+ }  \ ,
\end{eqnarray}
where $g=4$ takes account of valley and spin degeneracy.  
At $T=0$, the Fermi-Dirac distribution function is reduced to the Heaviside step function 
$f(\epsilon, \mu; T \rightarrow 0) = \eta_{+}(\mu - \epsilon)$,
In the long wavelength limit, the  real part of $\Pi^{0}(q,\omega)$ is  given by

\begin{eqnarray}
{\mbox Re}\  \, \Pi^{(0)}_{2D}(q_\parallel,\omega) &=& -\ \frac{  q_\parallel^2}{4 \pi \hbar^2 \omega^2}
\left\{ 4 \mu + \hbar \omega \ln \left(
\frac{2 \mu - \hbar \omega}{2 \mu + \hbar \omega} \right) \right\}
\nonumber\\
&-& \frac{  v_F^2 \, q_\parallel^4}{4 \pi} \left\{
\frac{3 \mu}{\hbar^2 \omega^4} - \frac{\mu \hbar^2}{4 \mu^2 - \hbar^2 \omega^2} +
\frac{1}{2 \hbar \omega^3}  \ln \left( \frac{2 \mu - \hbar \omega}{2 \mu + \hbar \omega} \right)
\right\} \ ,
\end{eqnarray}
and the   imaginary part by

\begin{equation}
{\mbox Im}\  \, \Pi^{(0)}_{2D}(q_\parallel,\omega) =     \frac{  q_\parallel^2}{4 \, \hbar \omega}
\left( 1 + \frac{1}{2} \frac{v^2 q_\parallel^2}{\omega^2}
\right)  \eta_+(2 \mu - \hbar \omega)  \ .  
\end{equation}
Upon substituting this form of the polarization function for the monolayer into
the integral equation for the inverse dielectric function, we have

\begin{equation}
K(z_1,z_2)= K_{SI}(z_1,z_2) - \sum_{i=1}^N
\Pi_{2D,i}^{(0)} (q_{||},\omega) \int_{-\infty}^\infty  dz^\prime\
K_{SI}(z_1,z^\prime) v(z^\prime-a_i) K(a_i,z_2)\ .
\label{eq:GG}
\end{equation}
We now set $z_1=a_i$ in Eq.\ (\ref{eq:GG}) and obtain

\begin{equation}
\sum_{j=1}^N \left\{ \delta_{jj^\prime} + \Pi_{2D;j}^{(0)} (q_\parallel,\omega)
\int_{-\infty}^\infty dz^\prime \  K_{SI} (a_{j^\prime},z^\prime) v(z^\prime-a_j)
\right\} \ K(a_j,z_2) = K_{SI} (a_{j^\prime}, z_2)  \ .
\end{equation}
These linear algebraic   equations may be solved simultaneously and their
solutions expressed in matrix form as

\begin{equation}
\left(
\begin{matrix} K(a_1,z_2)\cr
K(a_2,z_2)\cr
\cdots \cr
K(a_2,N_2)\cr
\end{matrix}
\right)= \frac{1}{S_c^{(N)}(q_{||},\omega)} \tensor{{\cal N}}^{(N)} (q_{||},\omega)
\left(
\begin{matrix} K_{SI}(a_1,z_2)\cr
K_{SI}(a_2,z_2)\cr
\cdots\cr
K_{SI}(a_N,z_2)\cr
\label{Melements}
\end{matrix}
\right)  \ ,
\end{equation}
where $S_c^{(N)}(q_\parallel,\omega)=\mbox{Det}
\tensor{{\cal M}}^{(N)} (q_\parallel,\omega)$ with
matrix elements given by

\begin{equation}
 {\cal M}^{(N)}_{jj^\prime } (q_\parallel,\omega)=
 \delta_{jj^\prime} + \Pi_{2D;j}^{(0)} (q_\parallel,\omega)
\int_{-\infty}^\infty dz^\prime \  K_{SI}(a_{j^\prime},z^\prime) v(z^\prime-a_j)
\end{equation}
and $\tensor{{\cal N}}^{(N)} (q_\parallel,\omega)$ is the transpose cofactor matrix of
$\tensor{{\cal M}}^{(N)} (q_\parallel,\omega)$.

\begin{figure}
\centering
\includegraphics[width=0.5\textwidth]{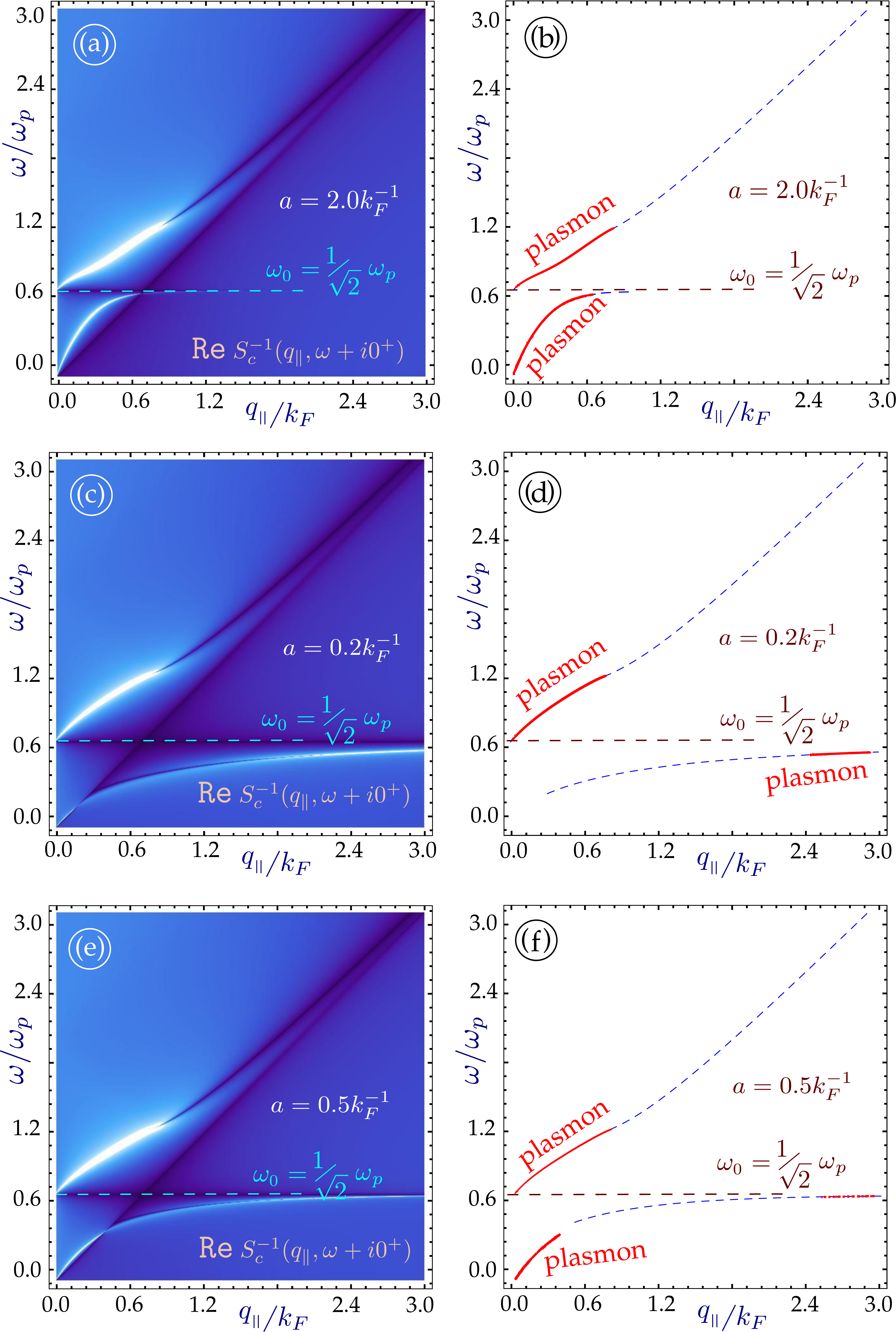}
\caption{(Color online)  Plasmon dispersion relation for a semi-infinite conductor  which
is Coulomb-coupled  to   monolayer graphene  for various surface-to-layer separations.
In the   panels $(a)$, $(c)$ and $(e)$ on the left-hand side, we present density plots of
 the inverse dispersion function $1/S_c^{N=1}(q_\parallel, \omega + i 0^{+})$ with
peaks corresponding to the undamped plasmon modes. The right panels $(b)$, $(d)$ and $(f)$
show  numerical solutions for the plasmon branches, both Landau damped and undamped.
The distances chosen are $a \, k_F = 0.2$, $1.0$ and $0.5$, correspondingly.
All  plots are provided for extrinsic graphene (doped) with zero energy bandgap.}
\label{FIG:1}
\end{figure}

In mean-field theory, we have  \cite{Kamen}

\begin{eqnarray}
\label{eqH21}
&& K_{SI}\left({\bf q}_\parallel, z, z^{\prime} ; \omega \right)
\nonumber\\
&=& \eta_+ (z) \Bigg\{\delta\left(z-z^{\prime}\right) -\frac{\varepsilon\left(q_\parallel\right)}
{1+\varepsilon\left(q_\parallel\right)} e^{-q_{\parallel}z} \delta(z^{\prime})
+    \frac{2\varepsilon\left(q_\parallel\right) }{1+\varepsilon\left(q_\parallel\right) }
K^{3D}_{\infty}\left({\bf q}_{\parallel}, z^{\prime} ; \omega \right) e^{-q_{\parallel} z}\eta_+ (-z^{\prime})
\Bigg\}
\nonumber\\
  &+& \eta_+(-z) \Bigg\{ v^{3D}_{\infty} \left({ \bf q}_{\parallel}, z ; \omega \right)
	\left( \frac{q_{\parallel} \varepsilon\left(q_\parallel\right) }{1+\varepsilon\left(q_\parallel\right) }
\delta(z^{\prime}) -\frac{2 q_{\parallel} \varepsilon\left(q_\parallel\right) }
	{1+\varepsilon\left(q_\parallel\right) }  K^{3D}_{\infty}\left({ \bf q}_{\parallel}, z^{\prime} ; \omega
	\right)\eta_{+} (-z^{\prime}) \right)
	\nonumber\\
  &+& \left[ K^{3D}_{\infty}\left({\bf q}_{\parallel}, z + z^{\prime} ; \omega \right)
	+K^{3D}_{\infty} \left({ q }, z - z^{\prime} ; \omega \right) \right] \eta_+ (-z^{\prime}) \Bigg\} \ ,
\end{eqnarray}
where

\begin{equation}
\label{eqH22}
\varepsilon\left(q_\parallel\right)^{-1}\equiv \frac{2q_{\parallel}}{\pi}
\int_0^{\infty} dq_z   \ \left[(q_{\parallel}^2+q_z^2)
\varepsilon_{\infty}^{3D} (\mathbf{q},\omega)\right]^{-1} \  ,
\end{equation}
and $\epsilon_{\infty}^{3D}(\mathbf{q},\omega)$ denotes  the three dimensional (3D)
 bulk dielectric function of the thick-slab material. Furthermore, Eq.\  (\ref{eqH21})
introduces the definitions

\begin{figure}
\centering
\includegraphics[width=0.5\textwidth]{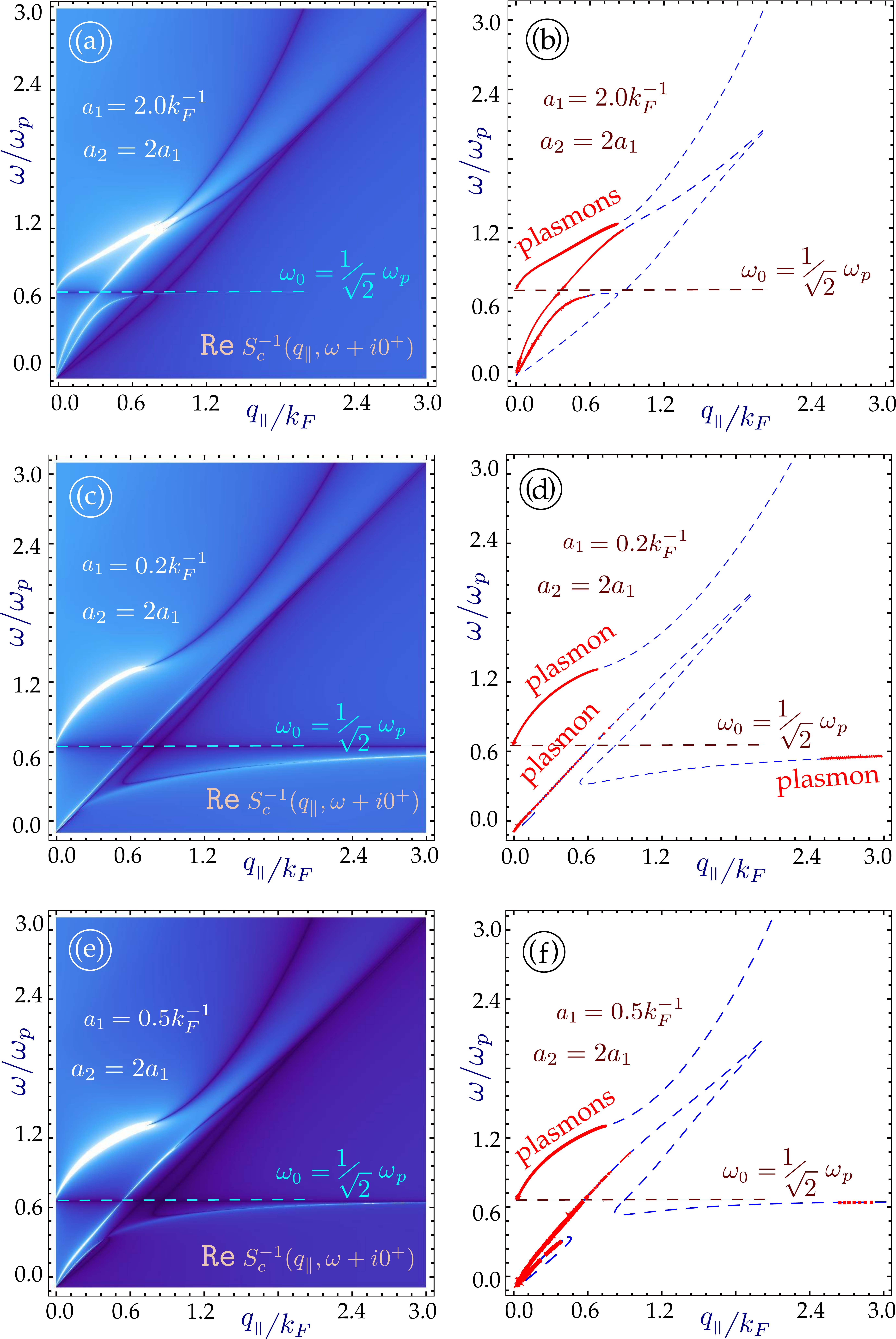}
\caption{(Color online)  Plasmon excitation spectra for a semi-infinite conductor
interacting  through the Coulomb interaction with   $N=2$ monolayer graphene  sheets
located at chosen distances from the surface. The left-hand panels
$(a)$, $(c)$ and $(e)$ give density plots of the inverse dispersion function
$1/S_c^{N=2}(q_\parallel, \omega + i 0^{+})$ with peaks corresponding to the plasmon
modes. The right panels $(b)$, $(d)$ and $(f)$ show the  numerical solutions for
the Landau damped (dashed blue lines)  and undamped (red curves)  plasmon branches. The plots show results for various
distances between the surface and the layers: $a_1 \, k_F = 2.0$, $0.2$ and $0.5$, respectively.
The second layer is placed at a distance   $a_2$, equal to  twice as large as $a_1$.
All the plots are provided for extrinsic graphene (doped) with
zero energy bandgap.}
\label{FIG:2}
\end{figure}

\begin{equation}
\label{eqH23}
K^{3D}_{\infty} \left(q_\parallel, z^{\prime}; \omega \right)
= \frac{1}{\pi} \int_0^{\infty} dq_z \ \frac{\cos q_z z^{\prime} }
{\varepsilon_{\infty}^{3D}(\mathbf{q},\omega)}
\end{equation}
and

\begin{equation}
\label{eqH24}
v_{\infty}^{3D}\left(q_\parallel, z^{\prime}; \omega \right)
= \frac{2}{\pi} \int_0^{\infty} dq_z \  \frac{\cos q_z z^{\prime}}{(q_\parallel^2 + q_{z}^2)
\varepsilon_{\infty}^{3D} (\mathbf{q},\omega) } \ .
\end{equation}
Making use of these results in Eq.\ \ (\ref{Melements}), we then obtain

\begin{eqnarray}
 {\cal M}^{(N)}_{jj^\prime } (q_\parallel,\omega)
 &=&   \delta_{jj^\prime}+  \frac{2\pi e^2}{\epsilon_s q_{||}}  \Pi_{2D;j}^{(0)} (q_\parallel,\omega)
\left [  e^{-q_\parallel\left| a_j-a_{j^\prime}  \right|}    +  e^{-q_\parallel( a_j+a_{j^\prime} )}
  \frac{\varepsilon(q_\parallel)}{1+\varepsilon(q_\parallel)} \right.
	\nonumber\\
	&+&\left.  \frac{2\varepsilon(q_\parallel)}{1+\varepsilon(q_\parallel)} e^{-q_\parallel a_j}
	\int_0^\infty  dz^\prime \ e^{-q_\parallel |z^\prime -a_j|  }
	K^{3D}_{\infty} \left({ q },   z^{\prime} ; \omega \right)  \right] \ .
\label{FULL}
\end{eqnarray}
If the bulk plasma within the semi-infinite slab is fully local in the sense that $\epsilon_{\infty}^{3D}(\mathbf{q},\omega) \to   \equiv \varepsilon_B(\omega)=1-\omega_p^2/\omega^2$,  in terms
of the bulk plasma frequency $\omega_p$,  then we use

\begin{eqnarray}
K^{3D}_{\infty} \left({ q },   z^{\prime} ; \omega \right) &=&   \frac{\delta(z^\prime)}{\varepsilon_B(\omega)}
\nonumber\\
v_\infty^{3D}   (q_\parallel,z^\prime;\omega)
&=& \frac{e^{-q_\parallel |z^\prime|}}{q_\parallel \varepsilon_B(\omega)}
\end{eqnarray}
from which  the corresponding local inverse dielectric function
$ K_{SI}^{local}\left({\bf q}_\parallel, z, z^{\prime} ; \omega \right)$ may be obtained using Eq.\ (\ref{eqH21}).

\begin{figure}
\centering
\includegraphics[width=0.95\textwidth]{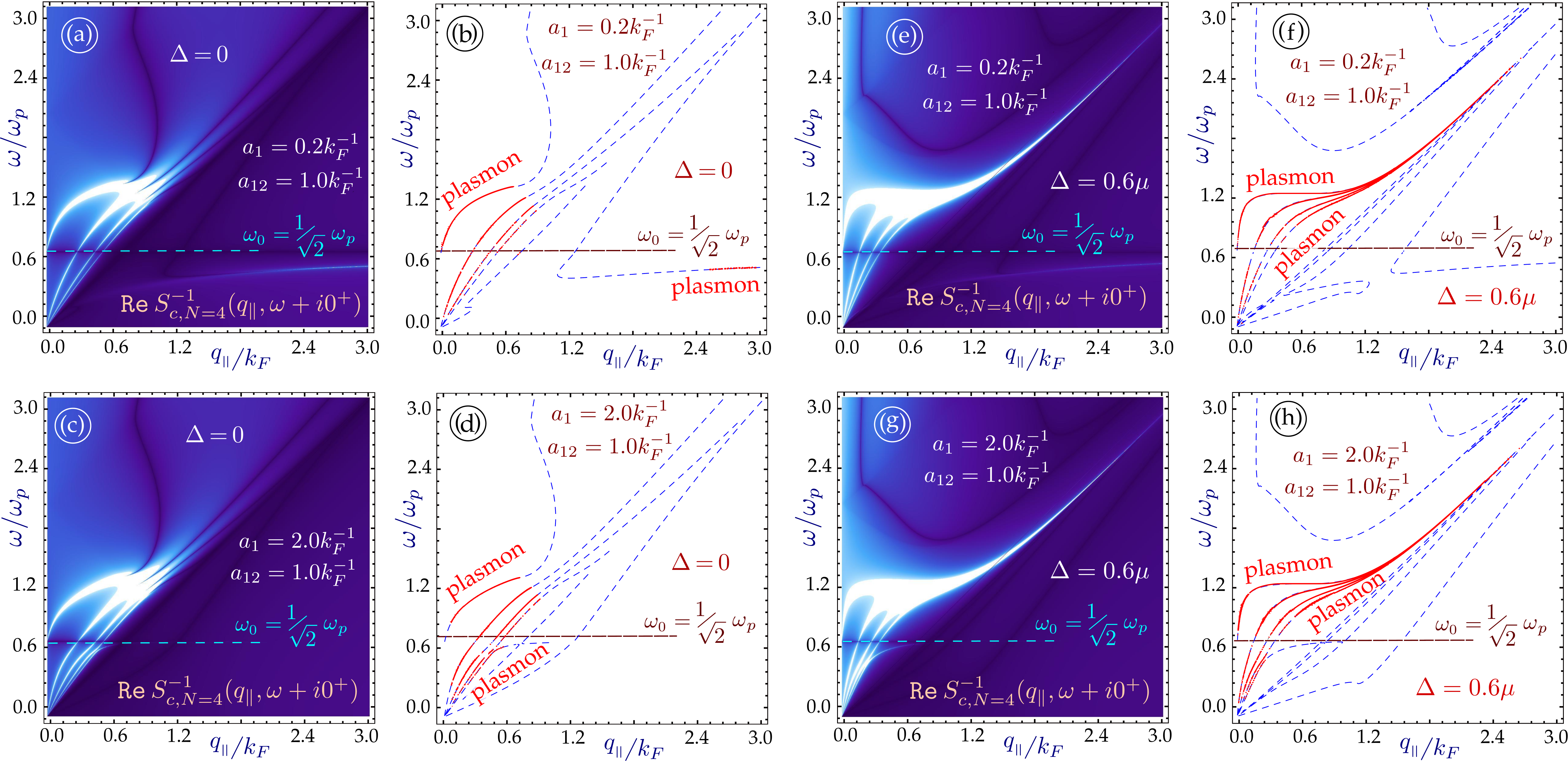}
\caption{(Color online) Density plots showing the bulk modes in the miniband
continuum and the line traces of plasmon  dispersion for $N=4$ graphene layers on a conducting substrate.
The line plots show damped (dashed blue lines)  and undamped (red curves) plasmon excitations.
In $(a)$-$(d)$, the layers are gapless, and in $(e)$-$(h)$  each layer has a gap $\Delta=0.6\ \mu$.
The layers are equally spaced with inter-layer spacing $a_{12}= k_F^{-1}$. The separation between
 the  first layer and the surface was chosen as  $a_1 \, k_F = 2.0$  $a_1 \, k_F =\mu/\hbar v_F$.
If the gap or number of layers is increased, the lowest branch does not re-appear for large
$q_\parallel$. It is more complicated than we thought.
}
\label{FIG:3}
\end{figure}

\medskip
\par

While Eq. (\ref{FULL}) is useful because of its apparent simplicity, it is necessary to
understand that its validity is restricted because the $q_z$-integrations in Eqs. (\ref{eqH22}
through \ref{eqH24}) extend over an infinite integration range. This blends the effects of
the boundary/image length scale with that of the  $q_z$-nonlocality dependence, eliminating
the possibility of an unrestricted limit $q_z\to 0$ and modifying the plasmon dependence
on $q_{\parallel}$. Additionally, the imaginary part of $\varepsilon_{\infty}^{3D}(\mathbf{q},\omega)$
is accounted for in these $q_z$-integrations, which consequently contributes  to    damping
of the surface plasmon modes even in the low-$q_{\parallel}$ limit.
The "nonlocal"   $q_{\parallel}$-correction to the surface plasmon, and its imaginary part
involving damping, depend  on the properties of the  bounding surface. But, there is a
range of applicability.  This can be seen by examining the parameter measuring the
importance of nonlocality in the bulk dielectric function  $\varepsilon_{\infty}^{3D}(\mathbf{q},\omega)$,
namely $p_c^M \sim \left(m^\ast \omega_p^2 /E_{thermal}\right)^{1/2}$, where the characteristic thermal
energy $E_{thermal}=\mu$ is the Fermi energy in the degenerate substrate with
electron effective mass $m^\ast$. For $q_\parallel <<p_c^M$, it is reasonable to neglect nonlocality,
at least in the surface plasmon frequency $\omega_s = \omega_p /\sqrt{2}$ as well as in comparison
with other more pertinent sources of nonlocal behavior (but it cannot be neglected in the
damping of the surface plasmon).


\section{Numerical Results}
\label{sec:3}

\par
\medskip
\par
\begin{figure}
\centering
\includegraphics[width=0.95\textwidth]{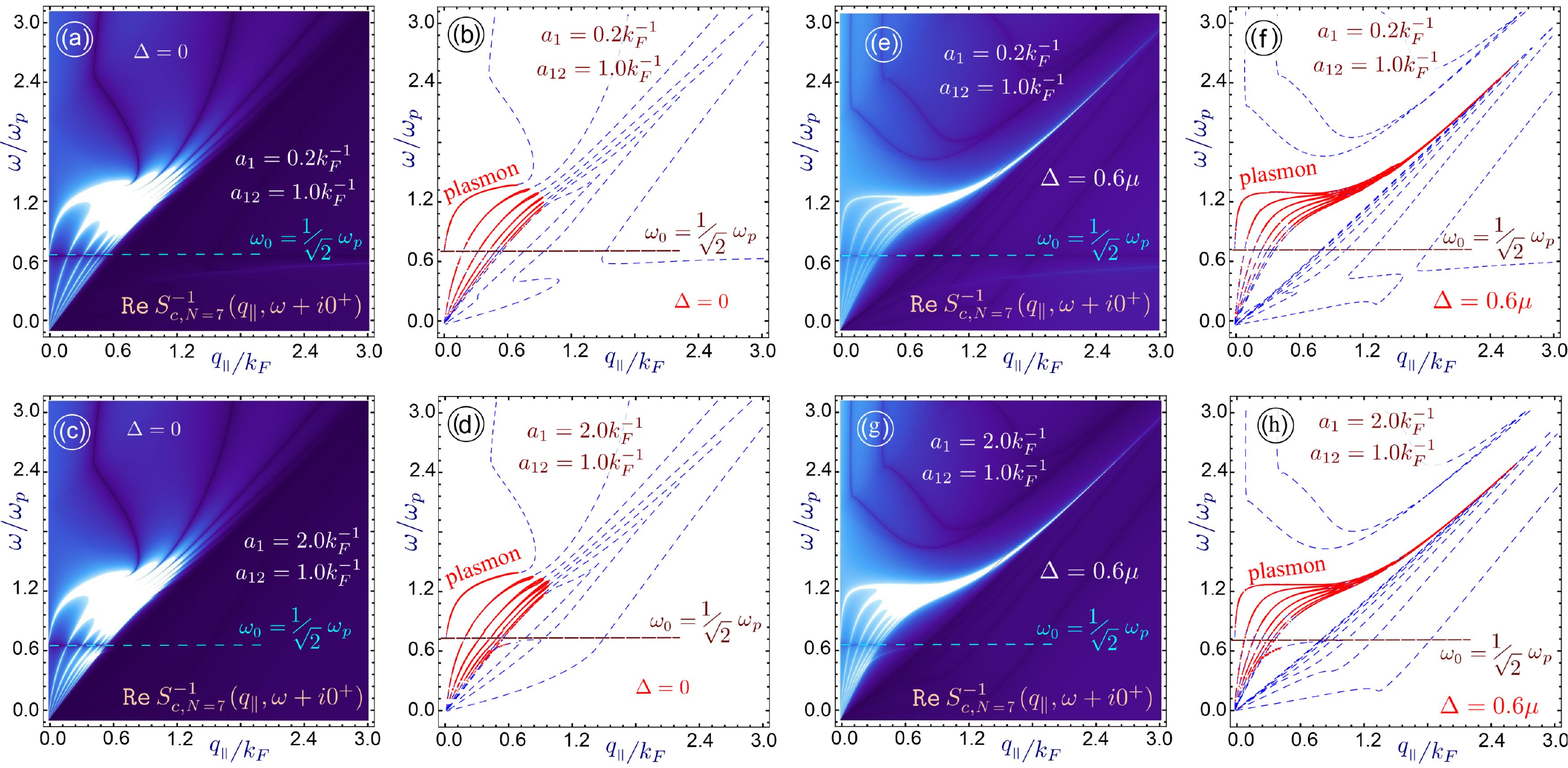}
\caption{(Color online) Plasmon dispersion for $N=7$ graphene layers on a conducting substrate.
As in Fig.\ \ref{FIG:3},   the layers  in $(a)$-$(d)$ are gapless, whereas in $(e)$-$(h)$  each layer has a
gap $\Delta=0.6\ \mu$.  The  inter-layer separation is $a_{12}=k_F^{-1}$. The first layer is at
distances  $a_1 \, k_F =  0.2$
from the surface of the semi-infinite conductor.
If the gap or number of layers is increased, the lowest branch does not re-appear for large
$q_\parallel$.
}
\label{FIG:4}
\end{figure}

In this section, we present numerical results for the plasmon dispersion  for a system consisting
of a semi-infinite conducting medium which is Coulomb coupled to $N=1,2,4$ layers  of graphene
as shown in  Figs.\ \ref{FIG:1}  through   \ref{FIG:4}. We note that both the plasmon solutions
and  damping by bulk modes in the miniband continuum crucially depend on the separation
between  the constituents as well as
the energy gap between the valence and conduction bands.   For a single layer, our results shown
in  Fig.\ \ref{FIG:1} demonstrate   that if the plasmon mode enters a region with
${\mbox Im}\   \Pi^{(0)}_{2D}(q_\parallel,\omega) \neq 0$, the  mode is  Landau damped. Our
calculations  also show  that when the distance $a$ is less than a critical value
$d_c \backsimeq 0.4 k_F^{-1}$,  in terms of the Fermi wave vector $\mu/\hbar v_F$,
 the lower acoustic plasmon mode is over-damped and
this behavior seems analogous to data reported experimentally \cite{Pol1,Pol2,Pol3,Pol4,1}. This
is obviously the case if the plasmon branch goes below the main diagonal
$\omega = v_F q_\parallel$. The damping, as well as the critical distance changes
in the presence of an energy bandgap for graphene.

\medskip
\par

Similar conclusions for a pair of graphene  layers electrostatically coupled to a semi-infinite
conducting material  are presented in Fig.\ref{FIG:2}. The principal difference between
the case when there are  two Coulomb-coupled layers is that if the distance of the layer
nearest the conductor  is less than  the critical separation $d_c$, both  symmetric and
antisymmetric    modes become damped, for  different  ranges of wave vector. We emphasize
that the upper plasmon branch (symmetric mode) remains almost unchanged for all cases,
either with one or two graphene layers.

\medskip
\par
The role played by the energy band gap is an important part of our investigation. For
monolayer graphene, an energy  gap leads to  an extended region of undamped plasmons
\cite{pavlop}. As we mentioned before, we pay particular attention to the regions outside of the
single-particle excitation continuum with $\texttt{Im} \Pi^{(0)}_{2D}(q_\parallel,\omega) = 0$,
since the plasmons in these regions are not Landau-damped.  In Figs.\ \ref{FIG:3}
and \ref{FIG:4}, we have plotted the plasmon dispersion relation for $N=4$ and $N=7$
graphene layers without and with an energy bandgap as well as for various distances
 between the nearest layer  to the conducting surface. These results show that for
 a conducting substrate surface plasmon frequency  denoted by $\omega_s=\omega_p/\sqrt{2}$,
 the surface plasmon frequency $\omega_c$ of the hybrid superlattice always lies below
 $\omega_s$. Furthermore, the  intensity of this mode depends on the distance
 of the  graphene layers from the surface of the conductor as well as the energy
 band gap between the valence and conduction bands of MLG. In the absence of a gap,
 our calculation shows that when the conductor surface-MLG separation exceeds a
 critical distance $d_c$, the intensity of the surface plasmon in the long wavelength regime
 is   high  and may be detected up to some cut-off wave vector $q_c$. For $q_\parallel>q_c$,
 the intensity of $\omega_c$ is very weak until the plasmon wave vector   exceeds some
 value $q_c^\prime$. However, when  the surface-MLG separation is less  than  $d_c$, the
 surface plasmon intensity in the long wavelength regime is weak and the mode only appears
 at shorter wavelength when $q_\parallel >q_c^\prime$. For gapped graphene, the surface
 plasmon frequency $\omega_c$    is completely suppressed when  the surface-layer separation
 is less than $d_c$.


\section{Concluding Remarks}
\label{sec:4}

The appearance of a surface plasmon polariton for a multi-layer structure
consisting of 2DELs was predicted  in the 1980's by Giuliani and
Quinn \cite{Giuliani} as well as by Jain and Allen \cite{Jain}.  This surface mode
is free from Landau damping or damping by bulk modes in the miniband continuum
and lies ``{\em above\/}" the continuum of
bulk modes.  Additionally, this surface mode only exists above a
critical wave vector $q_\parallel^\ast$ which is determined by the layer spacing
and the difference in the background dielectric constants   of the layers and
the surrounding medium.  When $q_\parallel^\ast$, there is damping by the bulk
plasmon modes.  As a matter of fact, the dispersion equation for the layered
superlattice structures investigated in \cite{Giuliani,Jain} is a special case
which may be obtained from our more general Eq.\ (\ref{FULL}) where we included the
effects arising from a substrate.

\medskip
\par

Very recently, in a series of experiments to determine the nature and behavior
of plasmon excitations in graphene interacting with metallic substrates,  Politano, et al.
\cite{Pol1,Pol2,Pol3,Pol4} showed how the dispersion and intensity may be affected in a substantial
way.  The experimental results show that self-doped graphene supported by a metal substrate  has two
  plasmon branches.  There is an acoustic plasmon, with   a linear dispersion,
and a nonlinear plasmon. Both plasmon branches   are similar in nature to those we
presented in Fig.\ \ref{FIG:1}, originating from the presence of a substrate.  The
present paper   investigating the effects of a substrate
on the plasmon excitations in a superlattice of graphene was stimulated by the
experimental results on supported graphene. We are aware of the theoretical work
on free-standing superlattice structures of MLGs \cite{Peeters1}, but the results there  do not
address the coupling to a metallic substrate which drastically
affect the plasmon dispersion relation as may be observed from Figs. \ \ref{FIG:1}
through \ref{FIG:2}.

\medskip
\par

The important conclusions of our work are as follows.   We formulated and exploited  a newly derived
expression for plasmon dispersion in a superlattice of 2D layers which are Coulomb-coupled to
a metallic substrate by taking into account the full nonlocality of the layers
as well as the underlying conductor.  We predict the existence of low-frequency nonlocal
plasmon excitations  at the vacuum-surface interface for various conditions of
surface-layer separation as well as the energy gap. When the separation between
the conducting surface and  the nearest layer is less than some critical distance $d_c$,
the surface plasmon may not exist in the long wavelength limit. We obtain a surface plasmon
at both intermediate as well as long wavelengths as this layer-surface separation is increased.
For this hybrid structure, the surface plasmon frequency lies below the surface
plasmon frequency for the semi-infinite substrate. Experimental verification of these
simulated results may be achieved using high-resolution electron-energy-loss spectroscopy
(EELS) \cite{Control}, for example.  This paper was inspired by recent experimental work
investigating the effect due to a metal on the collective plasmon mode of a single
layer of graphene \cite{Pol1,Pol2}. We presented a new approach for generating a
tunable  surface plasmon using    hybrid semiconductors. Additionally, our proposed
approach based on hybrid semiconductors can be generalized to include other novel
two-dimensional materials, such as hexagonal boron nitride, molybdenum disulfide and tungsten diselenide.

\acknowledgments
This research was supported by  contract \# FA 9453-13-1-0291 of
AFRL.

\newpage

%

%


\end{document}